\documentclass[usegraphicx,useAMS]{mn2e}
\usepackage{rotate}
\usepackage{times}
\newif\ifAMStwofonts
\AMStwofontstrue

\def\gs{\mathrel{\hbox{\rlap{\hbox{\lower4pt\hbox{$\sim$}}}\hbox{$>$}}}}
\def\ls{\mathrel{\hbox{\rlap{\hbox{\lower4pt\hbox{$\sim$}}}\hbox{$<$}}}}

\begin{document}

\title[Light bending and X--ray properties of accreting BH]
{A light bending model for the X--ray temporal and spectral properties
  of accreting black holes}
\author[G. Miniutti and A.C. Fabian ]
{G. Miniutti\thanks{E-mail: miniutti@ast.cam.ac.uk} and 
 A.C. Fabian \\ Institute of Astronomy, University of
  Cambridge, Madingley Road, Cambridge CB3 0HA 
}

\maketitle

\begin{abstract}
  { Some of the X--ray temporal and spectral properties of accreting
    black holes represent a challenge for current theoretical models.
    In particular, uncorrelated variability between direct continuum
    and reflection components (including the iron line, if present)
    has been reported in many cases.  Here, we explore a light bending
    model in which we assume a primary source of X--rays located close
    to a central, maximally rotating Kerr black hole and illuminating
    both the observer at infinity and the accretion disc.  We show
    that, due to strong light bending, the observed flux can vary by
    more than one order of magnitude as the height of the primary
    source above the accretion disc varies, even if its intrinsic
    luminosity is constant. We identify three different regimes in
    which the reflection--dominated component (and the iron line) is
    correlated, anti--correlated or almost independent with respect to
    the direct continuum. These regimes correspond to low, high and
    intermediate flux states of the X--ray source. As a general rule,
    the reflection component varies with much smaller amplitude than
    the continuum. X--ray observations of the Seyfert galaxy
    MCG--6-30-15 and of the Galactic black hole candidate
    XTE~J1650--500 reveal that a series of predictions of our model
    are actually observed: the consistent behaviour of the iron line
    flux and equivalent width (EW) with respect to the direct
    continuum, as well as the increase of the relative strength of
    disc reflection as the flux drops, all match very well our
    predictions. The iron line profile is predicted to be narrower in
    high flux states and broader in (reflection--dominated) low flux
    states, in fairly good agreement with observations of the
    best--studied case of MCG--6-30-15.  Observations of some other
    Narrow Line Seyfert 1 galaxies (e.g.  NGC~4051) also seem to
    support our model, which may explain what are otherwise puzzling
    characteristics of some sources. We also show that beaming 
    along the equatorial plane can enhance the re--emission
    of narrow reflection features from distant material during low
    flux states providing a possible contribution to the observed X--ray
    Baldwin effect.}
\end{abstract}


\section{Introduction}

The X--ray temporal and spectral properties of a class of active
galactic nuclei (AGN) show interesting behaviour that is difficult to
understand within the current theoretical view. This is most true for
those systems in which reflection spectral components and fluorescent
iron lines have been detected. Reflection in AGN is generally believed
to be associated with the reprocessing of the primary continuum by cold
material in the accretion disc close to the central black hole
\cite{gf91,mpp91} and/or by more distant material, such as the
putative torus of unified models \cite{anto93}.

The presence of broad and redshifted iron lines in some sources
indicates that special and general relativistic effects play an
important role in producing the line shape, supporting the idea that
reflection from the inner regions of an accretion disc is present
(Fabian et al. 1989; Laor 1991; Martocchia \& Matt 1996; Reynolds \&
Begelman 1997). However, the variability of the reflection component,
and most remarkably of the iron line, is not correlated in a trivial
manner to that of the observed continuum. The iron line does not
always respond to variations in the continuum as simple reflection
models predict. In some cases, an anti--correlation between the iron
line and the continuum has been reported; sometimes, the iron
line can appear to be constant while the continuum varies with large
amplitude (see e.g. Markowitz, Edelson \& Vaughan 2003).

Furthermore, the profile of the broad iron line (most remarkably in
the Seyfert 1 galaxy MCG--6-30-15) exhibits a singular behaviour:
qualitatively, the line tends to be very broad in low flux states,
while a narrower core is detected in high flux states
\cite{ietal96,wetal01,letal02}.  Moreover, in many sources, the line
equivalent width and the reflection fraction tend to anti--correlate
(or, in some cases to remain constant) with the continuum
\cite{lametal00,papaetal02}. If the continuum that we observe is the
same that illuminates the disc, these behaviours are difficult to
understand.

The uncorrelated variability between the iron line and the continuum
may be explained by requiring that it originates from a distant
reflector so that the variability of the illuminating continuum is
averaged out.  However, this interpretation conflicts with the
observation of broad and redshifted iron lines in some sources that
strongly suggests an origin close to the central black hole (e.g
Fabian et al. 2000; Reynolds \& Nowak 2003 for a review on iron
lines). Alternative explanations for the observed spectra which do not
require emission from the inner regions of the accretion disc have
been proposed (see e.g. Inoue \& Matsumoto 2003).

Here we consider a model based on the gravitational light bending
suffered by the radiation emitted in the near vicinity of a rotating
black hole with the aim of reconciling the observed puzzling
properties of some X--ray sources with the theory of reflection models
from accretion discs. We investigate the variability induced by light
bending by assuming that the primary source of hard X--rays is
centrally concentrated near the axis of a Kerr black hole. Variations
in the height of the primary source above the accretion disc produce
the bulk of the variability of both the observed continuum and the
reflection component \cite{mmk02,fv03}.  This idea was presented in
Miniutti et al. (2003) and successfully explained the puzzling
uncorrelated variability of the broad iron line and continuum seen in
a 325~ks {\it{XMM--Newton}} observation of the Seyfert 1 galaxy
MCG--6-30-15 \cite{fetal02,fv03,vfsub03}. After presenting the main
properties and predictions of the light bending model, we review the
phenomenology of some X--ray sources, and compare our predictions with
the available data.

\section{The light bending model}

In this section, we describe the most relevant assumptions and the
basic idea of our model for the spectral variability of accreting
black hole sources that exhibit a spectral disc reflection component
(see also Miniutti et al. 2003).

\subsection{Assumptions and computational setup}

We assume the presence of a central maximally rotating Kerr black hole
with specific angular momentum $a=0.998$. A geometrically thin
accretion disc lies in the hole equatorial plane and matter in the
disc is accreted along stable circular geodesics of the Kerr
spacetime. The disc extends down to the marginal stable orbit (with
radial coordinate $r_{\rm{in}} = r_{\rm{ms}} \simeq 1.24~r_g$) and has
an outer radius of $r_{\rm{out}} = 100~r_g$.  Relativistic effects
play a major role only in the near vicinity of the central black hole,
so that the inner disc radius is an important
parameter of the model, while the outer radius is not expected to have
significant effects on our results (see e.g. Martocchia, Karas \& Matt
2000). 

With this setup, it is clear that the spin parameter of the black
hole, which is here assumed maximal, is relevant for our results
because it fixes the value of the marginal stable orbit \cite{bpt}.
However, it is not yet completely clear that the inner disc radius
inferred from observations (e.g. via iron line diagnostics) completely
determines the black hole spin (e.g. Agol \& Krolik 2000; Krolik \&
Hawley 2002).  This is because, if emission from the plunging region
between the marginal stable orbit and the event horizon is considered,
even a moderate value of the black hole spin can reasonably well
reproduce the results that would be obtained neglecting the plunging
region emission and assuming a rapidly rotating Kerr black hole
(Reynolds \& Begelman 1997).
    
In this work we do not consider emission from matter accreting in the
plunging region at $r < r_{\rm{ms}}$. However, since we study the case
of a Kerr black hole with a disc extending down to the marginal stable
orbit,, the plunging region has a very limited radial extent ($\Delta
r = r_{\rm{ms}} - r_{\rm{hor}} \simeq 0.18~r_g$) so that the
contribution of these photons should be small. In fact, even in the
most extreme case we are considering here (a very low source height of
$h_s = 1~r_g$), the fraction of photons that illuminate the plunging
region is found to be negligible (less than 0.1 per cent of 
emitted photons).
    
As mentioned, the situation would be different in the case of a
non--rotating black hole in which case the plunging region radial
extent is $\Delta r = 4~r_g$ and might contribute to the disc emission,
producing results that would resemble those obtained in the maximally
rotating Kerr case (see e.g.  Reynolds \& Begelman 1997; Young, Ross
\& Fabian 1998). The analysis of the non--maximally rotating case, with the
inclusion of the plunging region emission, is potentially very
interesting, but beyond the scope of this paper which, for simplicity,
is restricted to the Kerr case.

The primary emission is due to a ring--like primary source of hard
X--rays that emits isotropically in its rest frame with a luminosity
described by a power--law. The primary source is located above the
accretion disc at a distance $\rho_s$ from the black hole axis and at
height $h_s$ above the equatorial plane. In this work, we restrict our
analysis to a centrally concentrated source with $\rho_s = 2~r_g$ and
height between 1 and 20 $r_g$. However, similar results are obtained if the
distance from the axis is kept within $\rho_s \approx 4~r_g$, while
the effects we shall discuss are reduced if $\rho_s$ is larger. We
also present some results for the case of a source on the rotation
axis, previously extensively studied by e.g Martocchia, Karas \& Matt
(2000).  It is not our purpose here to explore the full parameter
space of possible source positions. Our results are expected to be
qualitatively applicable in all the cases were a centrally
concentrated primary source of X--rays can be inferred from the
data (e.g. from the presence of a steep emissivity profile on the
disc, a broad iron emission line and/or an inner disc reflection
component).

The source could be physically realised by X--ray emission originating
above the very inner regions of the disc and related to magnetic
dissipation. In this case, dissipation of the black hole (or accreting
matter) rotational energy represents one possible physical energy
reservoir \cite{bz77,agolk00,li03}; the role of a non--zero black hole
spin is likely to be relevant and, in any case, dissipation is
expected to be concentrated in the very inner regions of the accretion
flow \cite{kI03,kII03}. The primary source could also be associated
with the inner part of an aborted jet producing relativistic particles
illuminating the disc (Malzac et al.  1998; Petrucci et al. in
preparation) or to the aborted jet itself \cite{ghise03}. Another
possible realisation could be self-Compton emission from the base of a
weak jet (see e.g.  Markoff, Falcke \& Fender 2001). Any other
mechanism producing a compact emission region above the innermost part
of the accretion flow would be compatible with our model; this kind of
geometry (often with a point--like source) has been previously adopted
in a number of works
\cite{mm96,hp97,ph97,reyetal98,baoetal98,luyu01,russ03}.  Since in many
reasonable cases a strong link between the accretion disc and the
primary source is likely (e.g. via magnetic fields), we assume that
the source is corotating with the same orbital velocity as the
underlying disc \cite{rus00}.
    
The orbital motion of the primary source justifies our choice of a
ring--like geometry for the emitting region instead of the more
commonly used point--like configuration: assuming for a moment that
the (corotating) primary source is point--like, the temporal
resolution of present observations (at least a few ks are needed to
extract meaningful spectra) is longer than the orbital timescale
close to the central supermassive black hole. Thus, our current
instruments can only detect spectral features from many orbital
revolutions, so that any information on the azimuthal position of the
corotating point--like source is lost in the data. In comparing
theoretical models with present observations, the primary source is
then best described by assuming a ring--like geometry. Moreover, we
cannot exclude that the primary source is itself axisymmetric and has
a real physical ring--like configuration if the associated physical
mechanism (e.g.  magnetic reconnection, hole rotational energy
extraction, link with a jet--like structure ...) had such a symmetry.

A fraction of the radiation emitted by the primary source directly
reaches the observer at infinity and constitutes the direct continuum
which is observed as the power--law component (PLC) of the spectrum.
The remaining radiation illuminates the accretion disc (or is lost
into the hole event horizon). The radiation that illuminates the disc
is reprocessed into iron fluorescent photons produced at 6.4~keV by
cold, non--ionised matter according to the work by George \& Fabian
(1991). In this work, we do not solve the radiation transfer within
the disc nor consider the effects of ionisation of the disc surface so
that the reflection continuum and possible shifts in the line
rest--frame energy are neglected.

Hereafter, the (neutral) iron line emission will represent the whole
reflection--dominated component (RDC) of the spectrum. We are of
course aware that a precise definition of the RDC should include the
reflection continuum and not only the iron line.  However, in this
work we are mainly interested in the variability properties and
correlations of the direct continuum (the PLC) and the reflection
components from the accretion disc and not in the detailed spectral
shape of these components. Since iron line emission and (disc) reflection
continuum are different aspects of the same physical phenomenon
(reprocessing of the illuminating flux), it is expected that they vary
together (and they must, within our model). Thus, we believe that, as
far as variability and correlation properties are concerned, the study
of the iron line provides an excellent approximation to the
variability of all the other components that are produced by
reprocessing of the illuminating continuum in the accretion disc, i.e.
the whole RDC.

We do not exclude that ionisation of the disc surface can contribute
to the RDC variability \cite{rfy99,naya00,naya01,coll03}. However, as
we shall discuss, many of the observed variability can be explained
even without invoking ionisation effects once special and general
relativity are properly taken into account.

\subsection{X--ray variability induced by gravitational light bending}

As already mentioned in the introduction, the variability of the iron
line (and of the RDC) in some sources is found to be uncorrelated with
the continuum, challenging theoretical models of reflection from
accretion discs. In the standard picture of reflection models, roughly
half of the photons emitted from the primary source reaches the observer
at infinity as direct continuum (or PLC), while the
remaining half illuminates the accretion disc, is reprocessed and
reaches the observer at infinity as the RDC of the spectrum. The
X--ray variability is assumed to be mainly due to intrinsic luminosity
variability of the primary source; the luminosity variation affects in
the same way both the direct continuum and the illuminating flux
on the accretion disc so that any RDC, including the iron line, has to
respond to the PLC variation. In this framework, a constant iron line
equivalent width and a constant value of the reflection fraction would
be produced, in contrast with many observations.

Here we adopt a different point of view with the aim of reconciling
the observed variability properties with theoretical models of disc
reflection. In this work, we assume that the intrinsic luminosity of
the primary source is constant and that the observed variability is
induced by gravitational light bending of the primary photons emitted
by the source. The basic idea is that the relevant parameter for the
variability of both the PLC and the illuminating continuum on the disc
(which drives the RDC variability) is the height of the primary source
above the accretion disc. The main difference with respect to the
standard picture is that changes in the height of the primary source
affect in different ways the direct and the illuminating continua so
that the RDC is not anymore expected to be strictly positively
correlated with the PLC. 

As an example, if the source height is small (of the order of few
gravitational radii) a large fraction of the emitted photons is bent
towards the disc by the strong gravitational field of the central
black hole enhancing the illuminating continuum and strongly reducing
the PLC at infinity, so that the spectrum is expected to be
reflection--dominated. If the source height increases, the
gravitational potential that photons have to overcome to reach
infinity is reduced, so that more photons reach infinity and the
observed PLC increases; in the same time, the illuminating continuum
on the disc is, in general, reduced (but depends also on the fraction
of photons that are lost into the black hole event horizon). Finally,
if the height is very large so that light bending is not very
effective, the standard picture of reflection models with
approximately half of the emitted photons being intercepted by the
disc and the remaining half reaching the observer as the PLC is
recovered.

It is clear that reflection responds to variations in the illuminating
continuum but not necessarily (or not in an intuitive way) to changes
in the direct, observed continuum that reaches the observer at
infinity as the PLC.  The illuminating continuum is the radiation
emitted from the primary source {\it{as it is seen from an observer on
    the accretion disc}}, while the direct continuum (or PLC) is the
same radiation but {\it{as it is seen from an observer at infinity}}.
The very different location (and thus the very different gravitational
field) and state of motion of the two observers make all the
difference between the illuminating and direct continuum.  The RDC
that we detect at infinity and its correlation properties with the PLC
have then to be computed from the illuminating continuum in a general
relativistic framework that takes into account all the effects on
photon propagation in the strong gravitational field of the central
(Kerr, in the present work) black hole.

It is also clear that our assumption that the intrinsic luminosity of
the source is constant is a simplification. In the general case, we
expect that the variability is due to both intrinsic variations and
light bending effects. However, the purpose of this work is to
present the model and to show that light bending, even alone, can
reproduce many of the observed spectral variability properties of some
X--ray sources.

\section{Predictions of the light bending model}
\begin{figure}
\rotatebox{270}{
\resizebox{!}{\columnwidth}
{\includegraphics{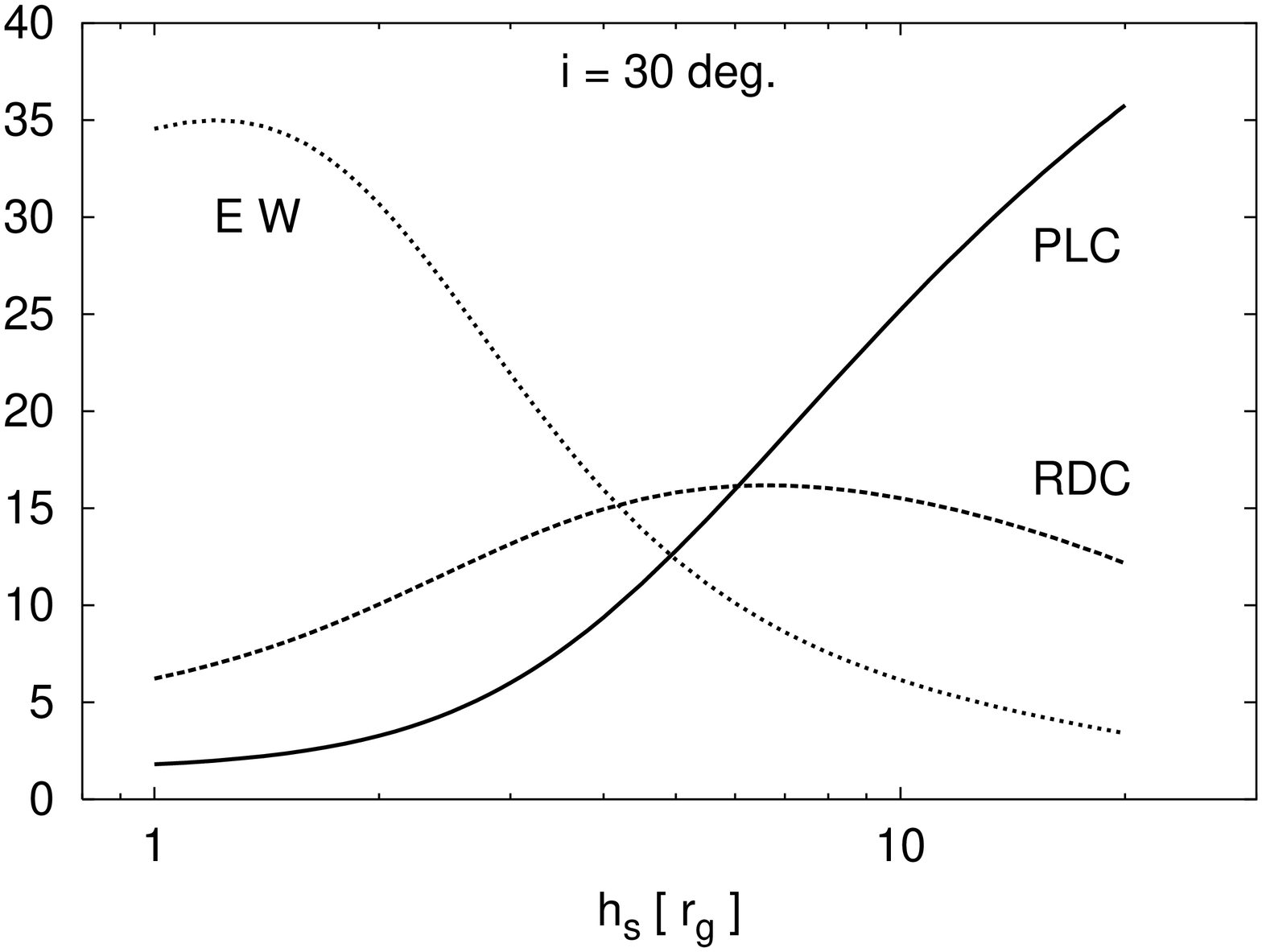}}}
\rotatebox{270}{
\resizebox{!}{\columnwidth}
{\includegraphics{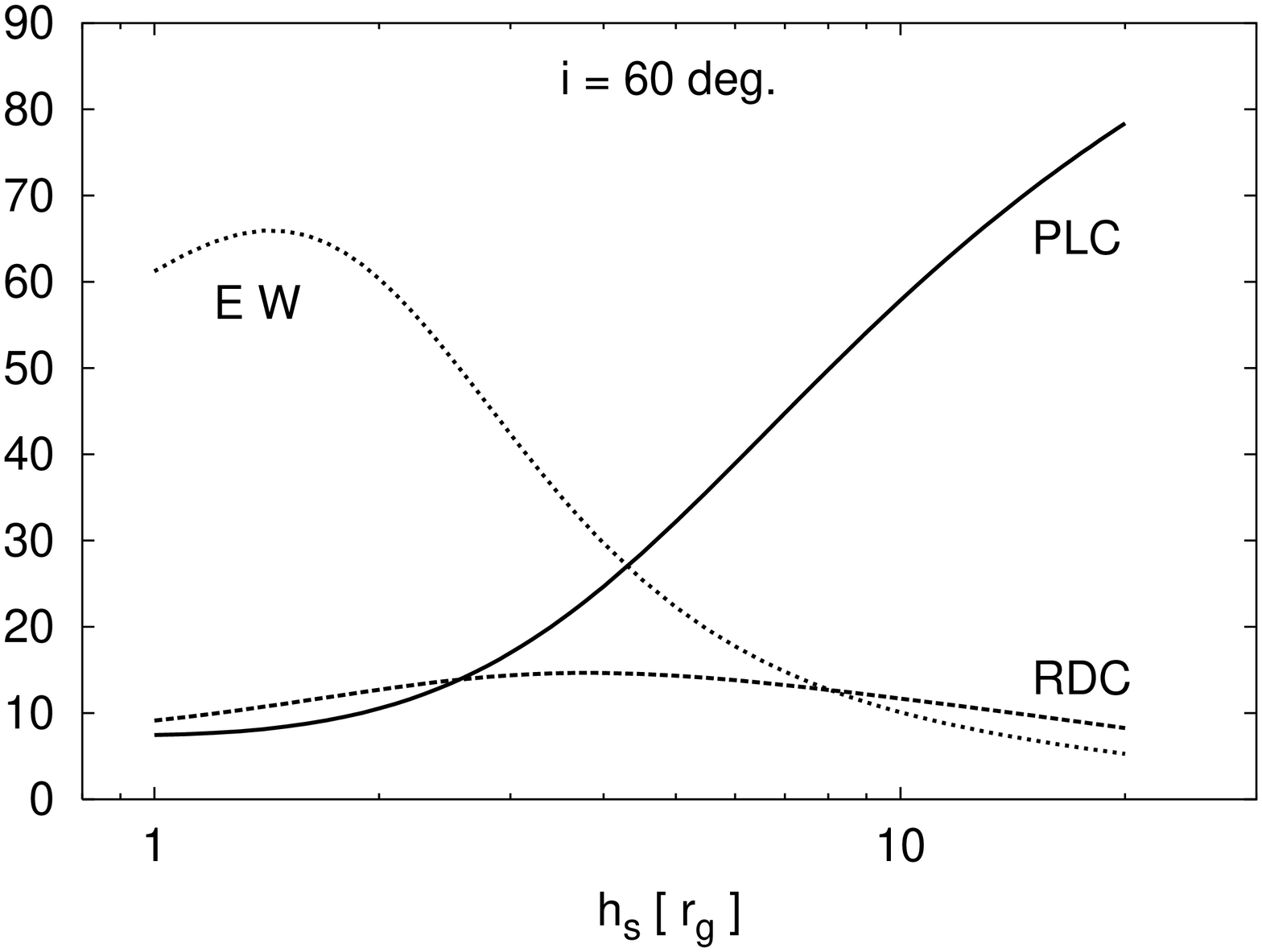}}}
%\vspace{0.3cm}
\caption{The iron line equivalent width (EW), the
  reflection--dominated component (RDC, represented by the iron line)
  and the direct continuum flux (PLC) as a function of the height
  $h_s$ of the primary source above the equatorial plane. The
  variations in the continuum and line flux are obtained by varying
  the source height from 1 to 20 $r_g$ at fixed intrinsic luminosity.
  The variability is then induced by light bending alone. The source
  is located at a distance of 2 $r_g$ from a Kerr black hole rotation
  axis and it is represented by a ring--like source corotating with
  the accretion flow. The top panel refers to an observer inclination
  of 30 degrees while the bottom panel is for an inclination of 60
  degrees. Units are arbitrary.}
\label{fh}
\end{figure}
\begin{figure}
\rotatebox{270}{
\resizebox{!}{\columnwidth}
{\includegraphics{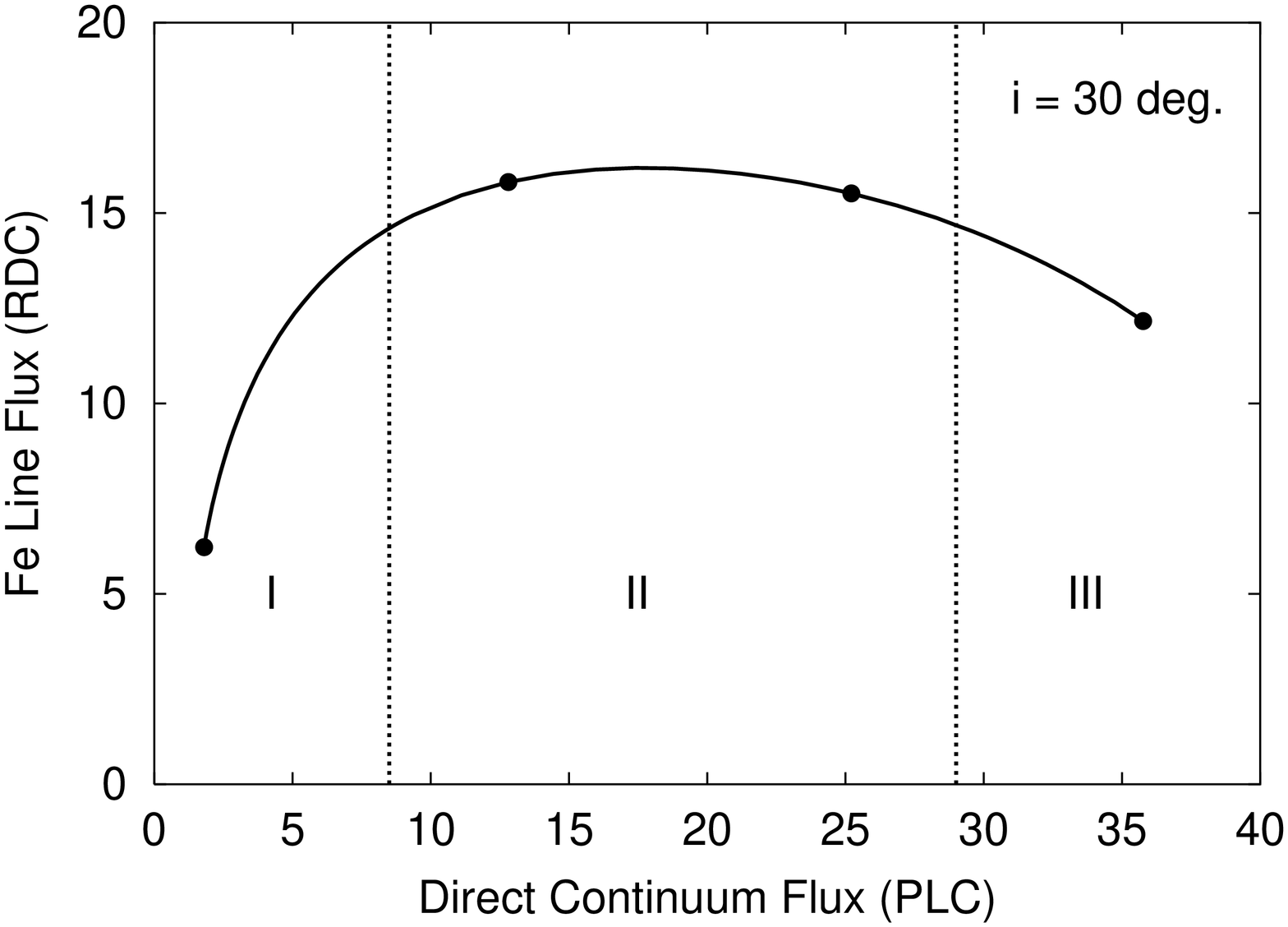}}}
\rotatebox{270}{
\resizebox{!}{\columnwidth}
{\includegraphics{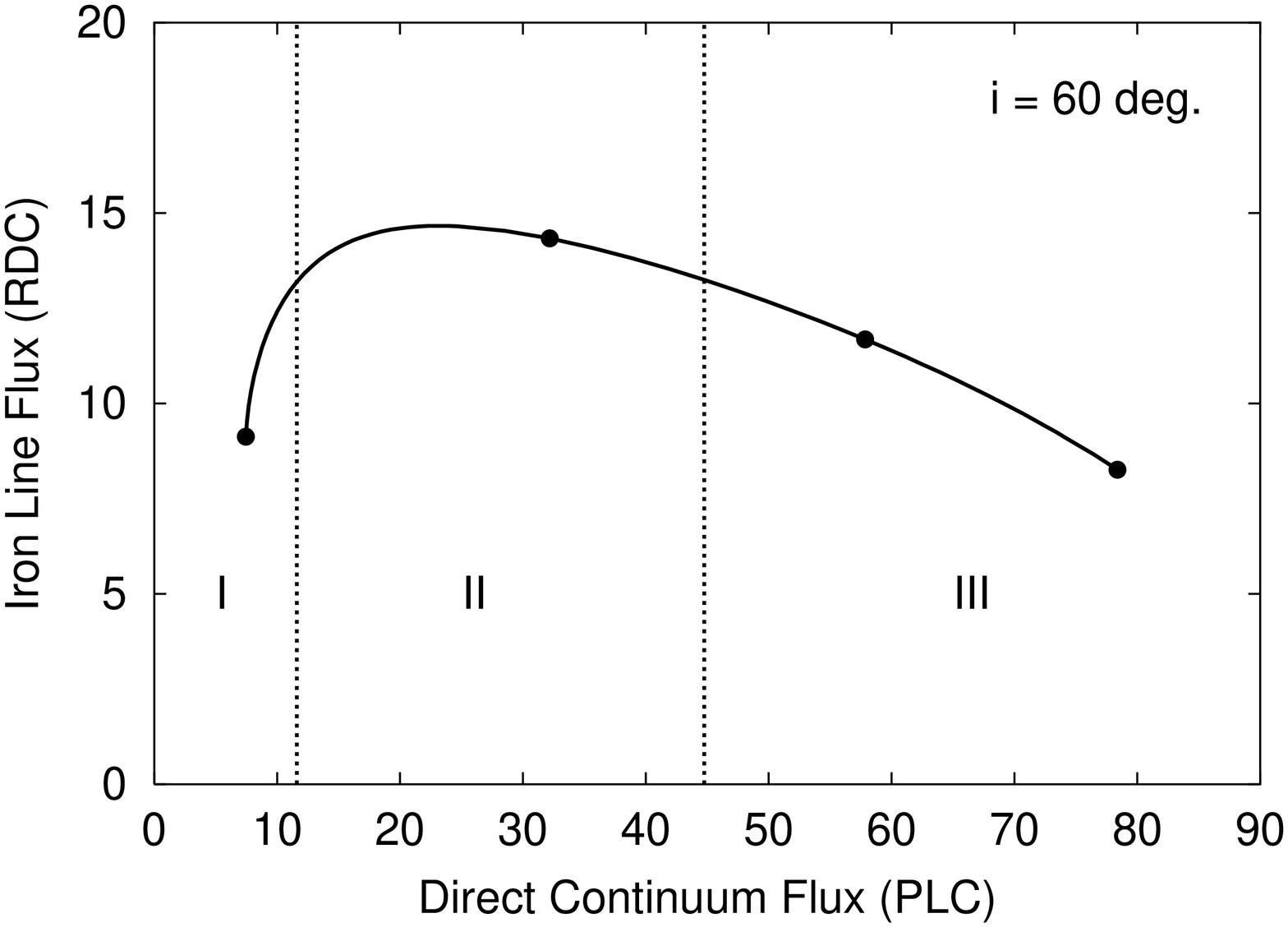}}}
%\vspace{0.3cm}
\caption{The iron line flux (RDC) as a function of the direct continuum
  flux (PLC) for an observer inclination of 30 (top panel) and 60
  (bottom panel) degrees. The source configuration is the same as in
  Fig. \ref{fh} and its height varies between 1 and 20 $r_g$. As a
  reference, black dots indicate different values of the source
  height: from left to right these are $h_s =
  1\,,\,5\,,\,10\,,\,20~r_g$. Three different regimes are identified
  (I, II and III, see text for details). Units are arbitrary.}
\label{ff}
\end{figure}

In Fig. \ref{fh}, we show the RDC (represented by the iron line) and
PLC fluxes and the iron line equivalent width (EW) as a function of
the height $h_s$ of the primary X--ray source for a corotating
ring--like source with $\rho_s = 2~r_g$ and $1\leq h_s \leq 20~r_g$. The
two panels refer to two different observer inclinations (30 and 60
degrees).  The luminosity of the primary source is assumed to be
constant so that the PLC and RDC variability is produced by light
bending alone. As shown in Fig. \ref{fh}, the PLC is strongly
correlated with the source height so that low flux states (low PLC)
are associated with low source heights and vice--versa. As the source
height changes, large variation of the PLC are allowed while the RDC
varies with much smaller amplitude. As an example, for an observer
inclination of 30 degrees, the maximum variation induced by a change
of the source height from 1 to 20 $r_g$ is about a factor 20 for the
PLC, while the iron line (and the RDC) varies at most by a factor 2.6.
At 60 degrees, one has a variation by a factor 10.5 for the PLC and
only 1.8 for the reflection component.

The iron line EW is almost always anti--correlated with $h_s$ (i.e.
with the direct continuum, or PLC, see Fig. \ref{fh}).  However, the
EW can be almost constant during extremely low flux states when the
source height is very low (see Fig.  \ref{fh}). As already mentioned,
one of the limitations of the present work is that we do not solve the
radiation transfer within the disc, so that we neglect the reflection
continuum.  Although this does not affect much the analysis of the RDC
variability (because the reflection continuum has to vary in the same
way as the iron line), the value of the iron line EW is affected. Our
results refer to the EW computed with respect to the power--law
continuum only, do not include the reflection continuum, and can be
tested against data if, during data analysis, the same procedure to
compute the line EW is adopted (see e.g.  Miniutti, Fabian \& Miller
2003). The main effect of the reflection continuum is expected to be a
saturation of the line EW in those situations where reflection
dominates the spectrum, i.e. when the source height is low (see e.g
Ross \& Fabian 1993; Ballantyne \& Ross 2002). Thus, a more realistic
picture for the EW behaviour is the one in which an asymptotic constant
value is reached as the source height decreases. We shall discuss again
this effect later in this Section.

\begin{figure}
\rotatebox{270}{
\resizebox{!}{\columnwidth}
{\includegraphics{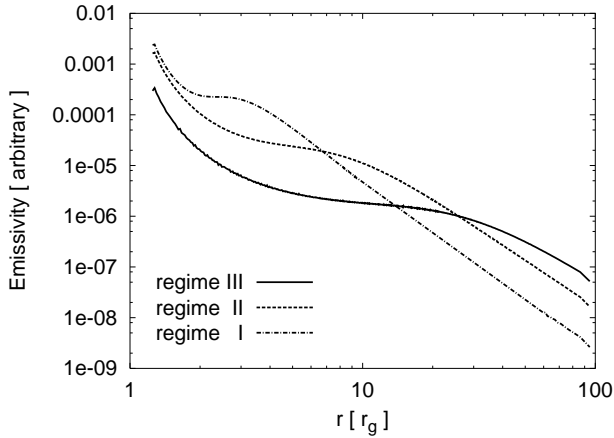}}}
%{\includegraphics{emireg.eps}}}
%\vspace{0.3cm}
\caption{We show the typical emissivity profiles on the accretion disc
  during regimes I, II, and III for the same source configuration as
  in the previous Figures. As a general rule, the emissivity is
  steeper in the inner regions of the disc and flatter in the outer.
  Moreover, the overall shape is steeper going from regime III (high
  source height and high PLC flux) to regime I (low source height and
  low PLC flux) because, as the source height is lower, the inner
  regions of the disc are more illuminated than the outer.}
\label{emi}
\end{figure}

\begin{figure}
\rotatebox{270}{
\resizebox{!}{\columnwidth}
{\includegraphics{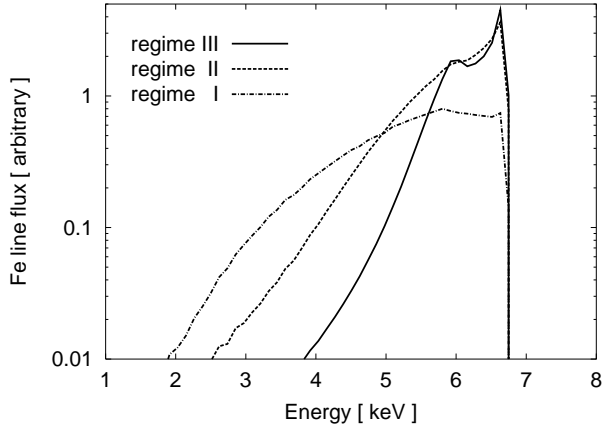}}}
%{\includegraphics{offreg.eps}}}
%\vspace{0.3cm}
\caption{We show the typical iron line profiles during regimes I, II,
  and III with the same primary source configuration as in the
  previous Figures, for an observer inclination of 30 degrees. A
  logarithmic scale is used in the y--axis to help to visualise the
  differences. The line is broader and fainter during regime I
  (low PLC flux states), reaches its maximum flux during regime II
  (intermediate PLC flux states) and becomes narrower and slightly
  fainter again during regime III (high PLC flux states), mainly
  because the red wing is reduced. As shown in Miniutti et al. (2003),
  the iron line profile of regime II provides a reasonably good
  description of the iron line of MCG--6-30-15 in the long
  {\it{XMM--Newton}} observation by Fabian et al. (2002a).}
\label{lines1}
\end{figure}

The relationship between the iron line flux (or RDC) and the direct
continuum that we observe (or PLC) is illustrated in Fig. \ref{ff}. We
show the iron line flux as a function of the observed direct continuum
for an inclination of 30 degrees (top panel) and 60 degrees (bottom
panel).  The plot of the RDC flux versus the direct continuum shown in
Fig. \ref{ff} allows to identify three regimes in which the behaviour
is clearly different (regimes I, II and III, corresponding to labels
in Fig. \ref{ff}). The typical emissivities on the disc and typical
iron line profiles obtained in the three different regimes are shown
in Fig.  \ref{emi} and \ref{lines1}.  The following discussion defines
the regimes, illustrate their properties and refers to Figures
\ref{fh}-\ref{lines1}, where the same primary source configuration has
been used.

\begin{enumerate}
\item {\emph{Regime I}}: in this regime the observed direct continuum
  is very low. Regime I corresponds to a low height of the primary
  source, where the strong light bending suffered by the primary
  radiation dramatically reduces the observed PLC at infinity (see
  Fig. \ref{fh} and \ref{ff}).  The iron line flux and the direct
  continuum are positively correlated. Since the flux illuminating the
  disc is concentrated in the inner regions, a steep emissivity
  profile and a broad and redshifted iron line are produced, as shown
  in Fig.  \ref{emi} and \ref{lines1}.  The iron line equivalent width
  is almost constant at very low heights and starts to decrease as the
  transition to regime II is approached. Because the RDC varies less
  than the PLC, the spectrum is more and more reflection--dominated as
  the source height and the direct flux are lower (see also Section
  3.1). The effect of the reflection continuum, that we neglected in
  this work, would be to produce an even more constant line EW in this
  regime. This is because, as the spectrum becomes
  reflection--dominated, the line and the continuum (almost only
  reflection) vary together so that the line EW remains constant
  reaching an asymptotic value.  We must then expect an almost
  constant line EW during regime I, corresponding to a low PLC flux
  and to a reflection--dominated spectrum. Regime I is associated
  with $h_s \ls 2-4 r_g$, depending on the observer inclination.
\item {\emph{Regime II}}: we identify this regime by requiring that
  the iron line flux varies less than 10 per cent around its peak
  value.  In other words, the iron line (and any RDC) has basically
  constant flux while the direct continuum (PLC) can vary by a factor
  $\sim$ 3--4 (see Fig. \ref{ff}).  The line EW is anti-correlated
  with the PLC (see Fig.  \ref{fh}) and changes in the line profile
  with flux are subtle and difficult to detect. We do not expect a
  large effect on the line EW caused by the reflection continuum in
  this regime, so that the anti--correlation will be maintained.
  Regime II was studied in more detail in Miniutti et al. (2003) and
  accounts for the variability of about a factor 4 in the PLC together
  with an almost constant iron line as revealed by {\it{ASCA}} and
  {\it{XMM-Newton}} long observations of MCG--6-30-15
  \cite{sif02,fv03}. During this regime, the RDC reaches a maximum.
  Regime II corresponds to $2-4 r_g \ls h_s \ls 7-13 r_g$,
  depending on the inclination.
\item {\emph{Regime III}}: in this regime light bending is less
  effective (but still present) and the observed direct continuum at
  infinity can be large. In regime III, the iron line flux is
  anti--correlated with the PLC and the line EW continues to decrease
  with the PLC, as in regime II. The line profile is much narrower
  than in regimes I and II because the emissivity profile on the disc
  is flatter, as a result of nearly isotropic illumination (see Fig.
  \ref{emi} and \ref{lines1}).  When the source height is very large
  (say $h_s \geq 30~r_g$, not shown in the Figures) and light bending
  does not affect the emission anymore, both the line and the direct
  continuum (and, of course, the EW) tend to their asymptotic values.
\end{enumerate}

These results are in good agreement with the analysis of the
long--term variability of a selection of AGN based on data from the
{\it{ Rossi X--ray Timing Explorer (RXTE)}} as reported by Papadakis
et al. (2002). The authors found a general anti--correlation between
the iron line EW and the continuum flux. The light bending model
predicts that the EW is most of the time anti--correlated with the
continuum, as shown in Fig. \ref{fh}. The only exceptions are during
low flux states of regime I or extremely high flux
states of a deep regime III where an almost constant EW can be
produced.  Moreover, the EW of the iron line is not necessarily
correlated with the reflection--dominated component flux. This is
especially true in regime I (almost constant EW while the RDC
increases with the continuum) and regime II (decreasing EW and almost
constant RDC). Only regime III is characterised by a clear positive
correlation between the EW and the RDC (both anti--correlated with the
continuum for $h_s$ smaller than about 30~$r_g$).

We point out again that, considering the three regimes altogether, the
iron line (and RDC) varies with much smaller amplitude than the PLC. This
behaviour is also observationally confirmed by the long--term
variability of seven Seyfert 1 galaxies studied in detail by
Markowitz, Edelson \& Vaughan (2003). 

As mentioned, it is also possible (and very likely) that intrinsic
luminosity variations of the primary source are superimposed on those
due to height changes and that they exist both on short and long
timescales.  This will lead to correlated variability between the PLC
and the RDC that may be most observable when the
source height undergoes a period of relative constancy (or, of course,
when light bending is negligible, say $h_s \geq 30~r_g$).

\subsection{Low--flux states and reflection--dominated spectra}
\label{reflectionfraction}

\begin{figure}
\rotatebox{270}{
\resizebox{!}{\columnwidth}
{\includegraphics{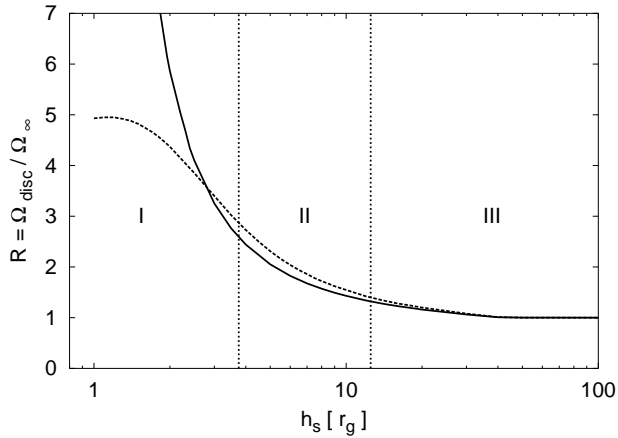}}}
%{\includegraphics{Rfrac.eps}}}
%\vspace{0.3cm}
\caption{ The reflection fraction R as a function of the height of the
  primary source above the equatorial plane. The asymptotic value
  (R$=$1 and $\Omega_{\rm{disc}} = \Omega_{\infty} = 2\pi$)
  corresponds to half of the primary source radiation reaching the
  disc and half infinity. The cases of a source on the rotation axis
  (solid) and of a corotating source at $2 r_g$ from the axis (dashed)
  are shown. We also show the three different regimes defined for an
  inclination of 30 degrees (appropriate for MCG--6-30-15 and
  NGC~4051). The variation of R with $h_s$ is due to light bending
  alone.}
\label{Rfrac}
\end{figure}

As the source height decreases, our model predicts that the spectrum
becomes more and more reflection--dominated because light bending
reduces the observed PLC flux and tends to enhance the illumination of
the disc and thus the RDC.  The reflection fraction is thus expected
to be large as a low flux state is reached.  Here we define the
reflection fraction to be $R \equiv \Omega_{\rm{disc}} /
\Omega_{\infty}$ i.e. the ratio between the solid angle subtended by
the primary source at the disc ($\Omega_{\rm{disc}}$) and at infinity
($\Omega_{\infty}$).

In Fig.  \ref{Rfrac} we plot the reflection fraction as a function of
the height above the disc for a source on the rotation axis of a
maximally rotating Kerr black hole and for a corotating source at $2
r_g$ from the rotation axis. We also show, as a reference, the three
different regimes defined above for an observer inclination of 30
degrees, which is appropriate for the cases of MCG--6-30-15 and
NGC~4051 that we shall discuss in detail later.

The maximum reflection fraction is about 5 for the off-axis source,
and tends to unity for large heights, where light bending effects are
negligible. Notice that the overall normalisation of R can vary from
source to source: here we assume that when light bending is negligible
(large $h_s$) half of the primary continuum illuminates the disc, the
remaining half reaching infinity.  For the on--axis case, we restrict
our plot in Fig.  \ref{Rfrac} to $R \leq 7$; if the height is lower
than about 2~$r_g$, as the source approaches the black hole event
horizon only a tiny fraction of the emitted photons are able to reach
infinity \cite{dl01,mmk02,mfgl03} and the reflection fraction can be
much larger. As an example, if the source is on--axis and $h_s =
1.25~r_g$, only about 1.2 per cent of the emitted photons reach
infinity, 21.9 per cent hit the disc (and are reprocessed as the RDC)
and 76.9 per cent are lost into the hole, so that the spectrum would
be interpreted as if the primary source had switched off, leaving only
a reflection--dominated component with $R \approx 18$.

The anisotropy ratio $\Omega_{\rm{disc}} / 2 \pi$ can be obtained from
R by considering that $\Omega_{\infty} \equiv 4\pi -
(\,\Omega_{\rm{disc}} + \Omega_{\rm{hole}}\,) \approx 4\pi -
\Omega_{\rm{disc}}$, where $\Omega_{\rm{hole}}$ is associated to
photons that are lost into the black hole event horizon. This is only
a very crude upper limit, as $\Omega_{\rm{hole}}$ should not be
neglected, especially in the case of low source height. However, since
the anisotropy ratio is sometimes used instead of R in both
observational and theoretical works, we give, as an example, the
approximate conversion $\Omega_{\rm{disc}} / 2 \pi \ls 2\,R /
(\,1+R\,)$ (see also Martocchia, Matt \& Karas 2002).

\subsection{Beaming in the equatorial plane}

There is another general/special relativistic effect that can produce
additional reflection spectral signatures. Due to gravitational lensing
and special relativistic beaming, the emission from the inner regions
of the disc is significantly beamed along the equatorial plane \cite{cunn75}.
Applications of relativistic beaming in the hole equatorial plane have
also been previously considered as a possible explanation for the
unusual extreme variability of IRAS~13224--3809 as well as for its
strong soft excess \cite{sunm89,boetal97}. 

\subsection*{Observing equatorial beaming}

\begin{figure}
\vspace{-0.9cm}
\hspace{-1.0cm}
\rotatebox{270}{
%\resizebox{!}{\columnwidth}
\resizebox{6cm}{1.1\columnwidth}
{\includegraphics{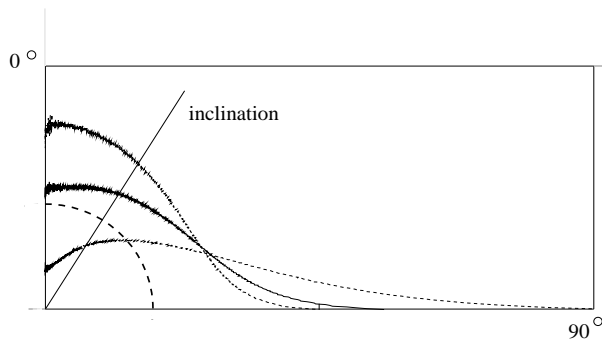}}}
%{\includegraphics{prova.eps}}}
\vspace{-1.0cm}
\caption{Polar diagram showing the number of photons per unit area that
  reach the observer at different inclinations. Photons are emitted by
  annuli in the equatorial plane of a Kerr black hole (the accretion
  disc). Looking at the left side of the plot, the emission radii
  ($r_s$) are 2, 4 and 8~$r_g$ from bottom to top. The dashed circle
  represents, as a reference, the case of isotropic emission. The
  intercepts of the lines with the y--axis is the number of photons
  (per unit area) detected at 0 degrees, while the intercept with the
  x--axis is the number of photons detected at 90 degrees. It is clear
  that photons are preferentially emitted along the equatorial plane
  so that the number of photons detected at 90 degrees is larger than
  at 0 degrees, contrasting strongly with the isotropic case.}
\label{beaming}
\end{figure}

The beaming in the equatorial plane of a Kerr black hole is
illustrated in Fig. \ref{beaming} where a polar diagram of the number
of photons (per unit area) that reach infinity being emitted from
three different rings in the equatorial plane ($\equiv$ accretion
disc) is shown. It is clear that the emission from the disc is far
from isotropic, especially if the emitting region is close to the
black hole resulting in a much larger number of photons detected at 90
degrees than at 0 degrees (see also Dabrowski et al. 1997 for a
similar figure). This effect is more pronounced for inner emitting
rings, as shown in Fig. \ref{beaming}. Thus, beaming along the
equatorial plane is more effective when the disc emissivity is
centrally concentrated, i.e. in low flux states (e.g. a regime I, see
Fig. \ref{emi}).

The beaming of the emission along the equatorial plane is maximal when
the emitting region lies in the inner equatorial plane of the black
hole (i.e.  radiation emitted/reflected from the inner regions of the
accretion disc).  However, the same effect arises also if the source
has a finite height above the disc (i.e. continuum radiation emitted
from the primary X--ray source). If the primary continuum originates
at low heights, even this radiation is preferentially emitted along
the equatorial plane, although the beaming is reduced as the height
increases. Thus, during low flux states both the primary continuum
and the disc emission are preferentially emitted along the equatorial
plane.  In the following we describe some of the observable
consequences of equatorial beaming:
\begin{enumerate}
\item a distant reflector (torus) and/or a disc that slightly diverges
  from the equatorial plane will intercept most of the radiation that
  is emitted from the innermost regions of the disc. The net effect is
  that narrow reflection features could be present in the spectra,
  associated with reflection from distant matter.
  In general, one must expect larger narrow features in low flux
  states, when the emissivity on the disc is concentrated in the inner
  regions so that beaming along the equatorial plane is maximal (and
  some contribution from the beamed primary continuum is also
  expected).  The narrow features, if present, will superimpose on the
  broad and redshifted ones that are produced in the inner disc making
  it difficult to distinguish between narrow and broad components,
  especially in short observations or faint sources;
\item the reflection spectrum produced by the inner disk surface has a
  strong soft X-ray/EUV component due to oxygen and lower Z elements
  being highly ionised, so making the albedo very high there, and to
  bremsstrahlung emission. Together with the intrinsic thermal UV
  radiation from within the disk, this is beamed along the plane of
  the disk as shown by Fig. \ref{beaming}. If the Balmer lines such as
  H$\beta$ are then produced by EUV irradiation of clouds close in
  height to the disk (but at distances of thousands of gravitational
  radii) and the systems are observed at low/moderate inclination,
  then the observed line widths will be small. The anisotropic nature
  of the irradiation of the broad-line clouds in such systems may
  therefore help to make them appear narrow. Peterson et al (2000)
  have argued that the optical lines in NGC~4051 can be explained by
  the broad-line region (BLR) being flattened and viewed at
  low/moderate inclination.  Our model causes the BLR to appear
  flattened due to the anisotropic irradiation;
\item in the standard unified model, Seyfert 2 galaxies are viewed at
  high inclination \cite{anto93}. This means that those for which the
  black hole has a high spin (so that the disc extends close to the
  horizon) should have a strong reflection component in the
  transmitted radiation. If the X-ray absorption due to the
  surrounding torus and other gas is strong, such strong reflection
  may only be evident in the Compton reflection bump at about 30~keV;
\item the companion star in stellar mass, black hole X-ray binaries
  samples the equatorial radiation. If the inner disc radius is small
  and the binary observed at moderate to low inclination, the
  equivalent width of the iron line due to reflection on the star
  could then be greater than expected. This could be relevant to
  observations of Cyg X-1 (e.g. Miller et al 2002c). A contribution
  from the outer region of a flared disc could also increase the
  narrow line equivalent width;
\item we also point out that, since several microquasars (e.g
  GRS~1915+105; Mirabel \& Rodriguez 1999) are viewed at high
  inclination, the reflection spectrum could be very strong due to
  equatorial beaming, especially during low flux states.
\end{enumerate}

From the discussion above, it appears that narrow features are
expected to be larger during low flux states. A natural prediction of
our model is that the EW of narrow lines from the outer regions of the
disc, the torus, and/or the BLR should anti--correlate with the flux
of the source.  This prediction holds for single sources whose flux
changes, but could have some relation with the observed X--ray Baldwin
effect (anti--correlation between lines EW and luminosity, Baldwin
1977), first reported for narrow iron lines by Iwasawa \& Taniguchi
(1993) and then confirmed by Nandra et al. (1997) and Page et al.
(2003).
 
\subsection{Variability timescales}

The model we are proposing must be able to produce the variability we
observe in some X--ray sources without invoking extreme conditions.
Light bending--induced variability is associated with changes in
source height: if these changes are produced by different active
regions appearing at different times and different heights, the
induced variability will depend on the physical mechanism that
activates the different emitting regions. On the other hand, if the
height changes are produced by the motion of a single (or dominant)
X--ray emitting region, the variability timescale is determined by the
vertical velocity of the source. Notice that, in the latter case, more
than one single timescale is expected, e.g. a shorter one due to
small--scale height fluctuations, and a longer due to large--scale
source motion.

A variability by a factor 20 in the observed flux can be obtained for
an inclination of 30 degrees if the source height changes from 1 to 20
$r_g$. For a $10^7~M_\odot$ black hole, this variability can
be produced in less than 2~ks by a single
primary source even if the motion is not highly relativistic (say
$0.05~c$). With the same setup, the shortest possible doubling time
would be less than 200~seconds.

We must point out that the most extreme variability in the
{\it{XMM--Newton}} observation of MCG--6-30-15 by Fabian et al.
(2002a) is a factor 4 in about 10~ks, well within the
limits of the the light bending model and allowing a single source as
slow as $\approx 2.5\times 10^{-2}~c$. Some Narrow Line Seyfert 1
galaxies and other sources show remarkably fast and large flux
variations \cite{boetal97,remetal91} which appear in some cases to
exceed the ``efficiency limit'' \cite{f79,braetal99} with doubling
times of few hundreds of seconds. These variations could however be a
consequence of the light bending effects discussed here and need not
to be intrinsic to the primary continuum.

In their analysis of the long--term variability of a sample of Seyfert
1 galaxies, Markowitz, Edelson \& Vaughan (2003) found (besides the
weaker variability of the iron line with respect to the continuum)
that the iron line, like the continuum, exhibits stronger variability
towards longer timescales. It is not our purpose to explain this
behaviour here, but merely note that if the variability is produced by
the (vertical) motion of a single primary X--ray source, we expect, in
general, that the strongest variability in both PLC and RDC will be
seen on longer rather than shorter timescales. Variability on short
timescales could be due to small fluctuations in the source height
(producing weak variability of the PLC and RDC), while on longer
timescales the primary source might cover large distances producing
larger variations in both components (see e.g. Fig. \ref{fh}).

In the case of a stellar--mass black hole, strong variability in the
PLC can be produced in a much shorter timescale than in AGN because
the distance that the (single) primary source must travel scales with
the black hole mass.  Assuming, as before, a velocity of the primary
source of $0.05~c$, the PLC can vary by a factor 20 in a timescale
$10^6$ times shorter for a $10~M_\odot$ black hole than for a typical
AGN with $10^7~M_\odot$, i.e. such variation can be reached in few
milliseconds (again for an observer inclination of 30 degrees).

Even more dramatic variations are produced for a more face--on
inclination and, clearly, if the primary source is moving more
relativistically. As already mentioned, the variability may also be
associated with the appearance of different active regions at
different heights. In this case, no source motion is implied and large
variations may be seen, in principle, on arbitrarily short timescales.

\section{Some applications to AGN}

In this Section, we discuss the phenomenology of the Seyfert galaxies
MCG--6-30-15, NGC~4051 and other Narrow Line Seyfert galaxies and we
compare the observational results with the predictions of our light
bending model. 

\subsection{MCG--6-30-15}

The spectrum and variability of the Seyfert 1 galaxy MCG--6-30-15 in
its normal flux states can be accounted for by a phenomenological
two--component model consisting of a variable power law representing
the direct continuum and an almost constant reflection--dominated
component which contains the broad iron emission line
\cite{mac98,sif02}. This model is also supported by the analysis of
the correlation between fluxes in different energy ranges using both
{\it{RXTE}} data \cite{tum03} and {\it{XMM--Newton}} observations
\cite{vfsub03}. The analysis reveals the existence of a soft component
varying in normalisation and an additional almost constant harder
spectral component. The interpretation of these components as the PLC
and the RDC is natural.  The two--component model explains also the
rms spectrum of the source and the correlation between spectral slope
and flux. A detailed discussion of the long 2001 {\it{XMM--Newton}}
observation by Fabian et al. (2002a) that is representative of the
normal flux state of MCG--6-30-15 is made by Vaughan \& Fabian (2003).

The lack of correlation between the two main spectral components
represents a challenge for the standard picture of reflection models
because, as already stressed, any reflection feature should respond to
the variations in the continuum, especially if it is produced in the
inner regions of the accretion disc as the reflection spectrum of
MCG--6-30-15 (and most remarkably the iron line shape) indicates.
However, as pointed out by Fabian \& Vaughan (2003), the
two--component model for the variability of this source could be
perhaps explained self--consistently by taking into account light bending
effects in the near vicinity of the black hole.

This suggestion was explored in detail by Miniutti et al. (2003) where
the application of the light bending model to the case of MCG--6-30-15
was presented for the first time. The relationship between the RDC and
the PLC shown in Fig. \ref{ff} demonstrate how an almost constant RDC
(and iron line) is obtained during regime II despite a variation of
the PLC by a factor $\sim$ 4, in agreement with the observations by
Shih, Iwasawa \& Fabian (2002), Fabian \& Vaughan (2003) and with the
discussion presented in Vaughan \& Fabian (2003). 

Notice that during the 2001 observation, the line (and RDC) is indeed
likely to vary within 25 per cent (with respect to a continuum
variation of about a factor 4). Thus, this observation could be
described by including a contribution from a regime I, in which the
line is slightly more variable than in a pure regime II. The regime I
contribution would be associated with the lowest flux epochs of the
observation.

The iron line profile obtained from our model if the primary source
height is confined within about 8~$r_g$ provides an adequate fit to
the 2001 {\it{XMM--Newton}} observation, supporting our interpretation
\cite{mfgl03}. The emissivity profile is observationally best
described as a broken power law, steeper in the inner disc than in the
outer, in agreement with the light bending model predictions (see Fig.
\ref{emi}). The iron line EW is observationally found to be
anti--correlated with the continuum during this same 2001
{\it{XMM-Newton}} observation \cite{vfsub03} and this is the behaviour
we found for the EW during regime II (see Fig.  \ref{fh}).
Furthermore, the observed reflection fraction ($R \gs 1.4$ in the
{\it{XMM--Newton}} data and $R \approx 2$ in the simultaneous high
energy {\it{BeppoSAX}} data) is also consistent with regime II, as
shown in Fig. \ref{Rfrac}.

Summarising, the Seyfert galaxy MCG--6-30-15 in its normal flux state,
represented here by the long 2001 {\it{XMM--Newton}} observation,
exhibits an almost constant iron line and RDC, a PLC varying by about
a factor 4 and an iron line EW which is anti--correlated with the
continuum.  All these observational properties, together with the iron
line profile and the relative strength of disc reflection, can be
reproduced by considering the source in regime II (with a possible
contribution from regime I during the lowest flux epochs).

As already discussed in Miniutti et al. (2003), the light bending
model predicts some changes in the profile of the iron line at
different flux states. The line will appear narrower when the PLC flux
is high and broader when dim; differences which are subtle and
difficult to detect in observations during a normal state of the
source (i.e. a regime II according to our interpretation). However, if
the PLC had a significantly lower (higher) flux, a broadening
(narrowing) of the iron line should be detectable (see Fig.
\ref{lines1}).

These predictions fit very well with the observations of MCG--6-30-15
in the so--called ``Deep Minimum'' state of the source identified by
{\it{ASCA}} in 1994 \cite{ietal96} and observed again in 2000 with
{\it{XMM--Newton}} \cite{wetal01} where a very broad iron line was
detected. We point out also that during the 2000 ``Deep Minimum'' the
reflection fraction is poorly constrained by the data and could be
larger than 2.5, supporting the idea that the spectrum may become
reflection--dominated in the very dim states (see Fig. \ref{Rfrac}).
Large values of the reflection fraction can be inferred from the
{\it{ASCA}} 1994 ``Deep Minimum'' observation as well.  Furthermore,
the iron line profile in this low flux state is very well reproduced
by assuming an illuminating source on the rotation axis at $h_s = 3
r_g$ \cite{mmk02}. Notice also that evidence for a narrower iron line when
the source flux is high has been reported \cite{ietal96,letal02} in
agreement with the predictions of the light bending model. In
particular, Iwasawa et al. (1996), showed how the Fe line profile is
dominated by the narrow component during a long bright flare, while,
during the ``Deep Minimum'', the broad red wing completely dominates:
these observed profile changes match very well our predictions. 

The 2000 ``Deep Minimum'' state of MCG--6-30-15 has been recently
analysed in detail (Reynolds et al.  2003).  One of the most
interesting results is that the behaviour of the iron line with
respect to the continuum appears to be different from that of the
normal state of the source; in the low flux state, the iron line is
found to be positively correlated with the continuum producing an
almost constant line EW within a variation of about a factor 2 of the
continuum.  These results, together with the increase of the
reflection fraction in low flux states of MCG--6-30-15, strongly
suggest that the ``Deep Minimum'' state is best described by a regime
I where the iron line is broader and correlated with the PLC (see Fig.
\ref{ff}) and the reflection fraction larger than in regime II. A deep
regime I is also characterised by an almost constant line EW (see Fig.
\ref{fh} and the associated discussion on the role of the reflection
continuum during regime I), so that even this observational result is
recovered.

By comparing the observational results on MCG--6-30-15 during its
normal and low flux states with the theoretical predictions of the
light bending model, it then appears that the main spectral and
temporal properties of the broad iron line can be recovered by
interpreting the normal flux state as regime II and the low flux (or
``Deep Minimum'') state as regime I and by considering the variability
to be mainly due to changes in the height of the primary X--ray source
above the disc. The behaviour of the the iron line flux and EW with
respect to the continuum match very well our predictions.  Moreover,
the iron line profile changes are also in  agreement with our
computations and the fact that the overall spectrum appears to be more
and more reflection--dominated as the source flux decreases, further
supports one of the main predictions of the light bending model.

\subsection{NGC 4051}

The Narrow Line Seyfert 1 galaxy NGC 4051 is a low luminosity AGN that
exhibits long timescale flux changes. The source was observed in an
extremely dim state in 1998 simultaneously by {\it{Beppo}}SAX and RXTE
\cite{getal98,uetal99}. The 1998 observations were interpreted by
assuming that the active nucleus had switched off, leaving a
reflection--dominated spectrum from distant material.  The reflection
spectrum is indeed consistent with reflection of a primary continuum
with much higher flux than that seen. The straightforward explanation
is that the reflector lies at large distances from the continuum
source (more than 150 light days from the inspection of the light
curve on long time scales, see Uttley et al.  1999) and keeps track of
the continuum emitted before the central source switched off.

The study of the long--term variability of NGC 4051 can be a powerful
tool for testing the model we are proposing. Lamer et al.  (2003)
reported on nearly 3 years of monitoring of NGC 4051 by {\it{
    RXTE}} (see also Uttley at al. 1998; 1999). These studies revealed
that the primary source did not suddenly switch off in May 1998. The
light curve evolution is more consistent with the source moving from a
highly variable high flux state to the very low state of 1998 through
an intermediate one. Furthermore, the long timescale variability is
associated with the primary continuum itself and not e.g. to varying
obscuration. These observations support the idea that the long
timescale variability is dominated by changes in one underlying
parameter (that may be the source height).

The variability analysis by Lamer et al. (2003) makes use of the idea
that a constant reflection component from a distant reflector is
always present and revealed by the 1998 ultra--dim state.  The authors
then subtracted this contribution in the analysis during the other,
higher flux states. As a general rule, a positive correlation between
the broad (i.e. emitted from the disc) iron line and the direct
continuum is found, although epochs in which the line appears to
be constant while the continuum varies by a factor $\approx$~2 are
also observed.  If light bending is providing the most important
mechanism for the observed variability, these behaviours would
correspond to regime I and II respectively (see Fig. \ref{ff}). The line
equivalent width is found to be roughly constant or decreasing with
the continuum, consistently with regime I where the EW is almost
constant and regime II where it decreases with the continuum flux as
shown in Fig. \ref{fh}.

Recently, a {\it{Chandra}} observation of NGC 4051 during a 2001
low--flux state (not as low as the 1998 one) has been reported by
Uttley et al. (2003a).  Spectral curvature is present in the data and
the hard residuals (with respect to a simple power law description of
the continuum) cannot be explained in terms of reflection from distant
material.  Furthermore, the rapid broadband spectral variability is
also inconsistent with this hypothesis. The presence of some distant
reflector is not ruled out, but it does not contribute too significantly
to the hard continuum spectrum. This result does not necessarily
conflict with the interpretation of the 1998 spectrum because the 1998
continuum was fainter than the 2001 one and reflection from distant
material could have been dominating.

As pointed out in Uttley et al. (2003a), a possible explanation of the
hard spectral shape of the 2001 {\it{Chandra}} spectrum is that it is
associated with strong reflection from the inner regions of the
accretion disc. A good fit to the 3--10 keV data is obtained by
considering strong disc reflection (with a reflection fraction of the
order of 3 but not very well constrained) together with an associated
iron line produced in the very inner regions of the accretion disc
around a maximally rotating Kerr black hole. Results from a 2002
{\it{XMM--Newton}} observation during a similar flux state have been
recently reported by Uttley et al. (2003b): a broad iron line is not
formally required to describe the data (see also Pounds et al. 2003),
although some amount of disc reflection is probably present together
with reflection from distant material.

The 2001 and 2002 low state of NGC 4051, as observed by {\it{Chandra}}
and {\it{XMM--Newton}} and the 1998 ``ultra--dim'' state, provide a
possible test of the light bending model that was able to explain the
variability of MCG--6-30-15 \cite{mfgl03}. In this model, the low
states observed in NGC 4051 should correspond to a period during which
the primary X--ray source was located at low heights above the disc
(regime I according to the discussion in the previous Section). If
this is the case, the observed spectrum becomes naturally
reflection--dominated as most of the primary photons are bent onto the
accretion disc enhancing the reflection features with respect to the
primary continuum.

A reflection fraction of the order of that indicated in the analysis
by Uttley et al. (2003a) of the 2001 {\it{Chandra}} data can be
obtained if the primary source height is about $3~r_g$, thus in a
regime I (see Fig. \ref{Rfrac}). At these heights, the strong,
redshifted, broad iron line invoked by Uttley et al.  (2003a) as a
possible explanation of the hard residuals is naturally produced.  As
already mentioned, a broad line is not formally required to describe
the 2002 {\it{XMM--Newton}} data and the fainter 1998 {\it{BeppoSAX}}
and {\it{RXTE}} observations, which seems to rule out the light
bending model as a possible explanation of the variability in this
source.

However, NGC~4051 might represent a case where light bending effects
are extreme: the fact that NGC~4051 reached a state that can be seen
as if the primary source had switched off might suggest that the
primary source is even more centrally concentrated than in
MCG--6-30-15; indeed, if the source is low and on--axis, only a tiny
fraction of the emitted photons can reach infinity because of the
vicinity with the black hole event horizon, as discussed in Section
3.1. The spectrum is then completely dominated by reflection and
the most straightforward interpretation would be that the source had
switched off, as in the 1998 observations.

\begin{figure}
\rotatebox{270}{
\resizebox{!}{\columnwidth}
{\includegraphics{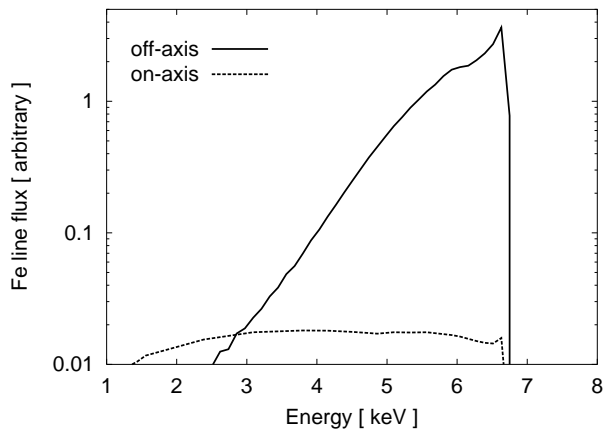}}}
%{\includegraphics{offon.eps}}}
%\vspace{0.3cm}
\caption{The iron line profiles for the cases of a primary source on
  the black hole rotation axis (on--axis) and $h_s = 1.7 r_g$, and for
  a source at $\rho_s = 2 r_g$ from the axis (off--axis) and $h_s = 6
  r_g$.  Both line profiles are computed for an inclination of 30
  degrees, appropriate for both MCG--6-30-15 and NGC~4051.  The
  off--axis case is shown for comparison and is a typical profile for
  the regime II: this kind of profile fits well the MCG--6-30-15 data
  (see Fabian et al. 2002a; Miniutti et al.  2003) and is thus
  detectable with {\it{XMM--Newton}}. On the other hand, the detection
  of the line relative to the on-axis case would be much more
  challenging, if not impossible, once the line is ``hidden'' in the
  associated reflection continuum.}
\label{lines2}
\end{figure}

We point out here that, in this case, the iron line would be so broad
that chances of detection would be seriously, if not totally,
compromised. The line profile would be basically featureless,
resulting in some spectral curvature below about 7~keV (depending on
the ionisation and inclination) rather than in a clear emission--like
feature. This is illustrated in Fig. \ref{lines2} where we compare the
iron line profiles obtained if an on--axis source at $h_s = 1.7~r_g$
is considered, with a typical regime II line profile that was able
to describe the 2001 {\it{XMM--Newton}} observation of MCG--6-30-15
and is thus detectable with this X--ray observatory: units are arbitrary but
common between the two computations in order to allow direct
comparison.

It is clear that the line resulting from the illumination by a very
low--height on--axis source will be much more difficult, if not
impossible, to detect against the associated disc reflection continuum
(and the other components from distant material that are likely to be
present in NGC~4051), possibly reconciling the light bending model
even with the 1998 observations of NGC~4051 where no evidence for a
broad line was reported \cite{getal98,uetal99}. We do not show the
line profile for a primary source at lower heights because it would be
almost invisible in the Figure, even using the logarithmic scale. The
fact that in the slightly higher flux states observed in 2001 with
{\it{Chandra}} and 2002 with {\it{XMM--Newton}} the presence of the
broad line is difficult to assess (Uttley et al.  2003a) or even not
formally required (Uttley et al. 2003b), while disc reflection is
probably present, might support our interpretation: if the height of
the on-axis source is slightly larger than in Fig.  \ref{lines2}, the
resulting line profile would remain almost featureless but would have
a slightly larger flux making possible, but difficult, its detection.

Thus we propose a possible scenario: the iron line would be detectable
when it results in a relatively prominent emission--like feature
during normal/high flux states of the source, while it would challenge
detection in low states (source height around 3~$r_g$ and
reflection fraction R of about 3) and completely avoid observability
in ultra--dim states ($h_s < 2~r_g$ and larger R) because of its
weakness and featureless spectral shape. 

Furthermore, the 1998 spectrum of NGC 4051 was interpreted as
dominated by reflection from distant material mainly because the
measured average 2--10 keV flux is more than one order of magnitude
lower than in any ``normal state'' while the iron line flux is only a
factor 3 lower than in previous observations (see e.g. Guainazzi et
al. 1996 and 1998). This interpretation seems the most natural and, as
explained, is not necessarily in contrast with the 2001 and 2002
observations. However, this extreme difference in the continuum and
line variations can be reached also by considering reflection from the
inner regions of the disc plus light bending effects. Light bending
alone can cause a large drop of the continuum flux as the source
height goes to zero together with a much smaller line flux variation
(see Fig. \ref{ff} and associated discussion). Moreover, the presence
of an additional component from distant material  can contribute to
further confuse the picture.

We point out that the radial ionisation structure on the accretion
disc, which we have neglected for simplicity, can play an important
role especially during a deep regime I where one can reasonably expect
the ionisation of the disc surface to be concentrated in the inner
regions of the disc where the illumination flux is much larger
\cite{gmfr93,gmfr96,rf93,br02}. Ionisation and/or thermal
instabilities \cite{naya00} could thus prevent or reduce the emission
of the red wing of the iron line, possibly changing its correlation
properties as well, and making even more difficult to detect the line
in very low flux states (see also Uttley et al.  2003b for a discussion of
this same possibility in the recently analysed 2002 {\it{XMM--Newton}}
observation of NGC~4051).

\subsection{Reflection--dominated spectra of NLS1 galaxies ?}

Some Narrow Line Seyfert 1 galaxies (NLS1) show similar properties
suggesting that their spectra may well be in some cases
reflection--dominated, although this interpretation is not unique.
Here, we focus on three sources, namely 1H~0707--495, IRAS~13224--3809
and IRAS~13349+2438. 

\subsection*{1H~0707--495}

1H~0707--495 was observed by {\it{XMM--Newton}} in 2000 October and
the results of the spectral analysis reported by Boller et al. (2002).
The most interesting result of this observation is the discovery a
spectral feature around 7 keV where a sharp drop of more than a factor
2 in the spectrum has been detected. One possible explanation of such
a drop is a partial--covering model provided by a patchy absorber
\cite{hetal80}. The spectrum can be modelled by a power--law with
three absorbers resulting in a very good fit
($\chi^2_\nu = 177 / 178$).

Alternative explanations to the spectral shape of 1H~0707--495 have
been proposed mainly by requiring the spectrum of the source to be
reflection--dominated \cite{betal02,fbetal02}. The drop at 7 keV may
be seen as the blue wing of a strong, broad iron K$_\alpha$ line. A
relativistically broadened line (with rest energy 6.4 keV and EW
of about 5~keV) provides a good fit to the data if the emission comes
from the inner regions of the disc around a maximally rotating Kerr
black hole. 

\subsection*{IRAS 13224--3809}

IRAS 13224--3809 was observed with {\it{XMM--Newton}} on 2002 January
19 \cite{betal03}. As in the case of 1H~0707--495, a sharp and
deep spectral drop was detected at about 8 keV. The observed spectral
features can be explained in terms of absorption if, once again, a
partial--covering model is invoked. 

An alternative is again provided if one interprets the drop as the
blue wing of a strong and relativistically broadened iron line
produced in the accretion disc (seen at about 60 degrees) around a
Kerr black hole \cite{betal03}; as in the case of 1H~0707--495, the
required EW is large (larger than about 5~keV).

\subsection*{IRAS~13349+2438}

IRAS~13349+2438 was observed by {\it{XMM--Newton}} on 2000 June 19 and
20 \cite{letal03}. A broad feature in the 5-7 keV region of the
spectrum is observed, together with an absorption edge at about 7.4
keV. A partial--covering model with the associated transmission edge
can describe the data but leaves residuals in the 5-7~keV band.

Another interpretation invokes reflection models which have the
advantage to explain self--consistently the edge and the broad 5-7 keV
feature. The best fit to the data is obtained with a phenomenological
model comprising ionised reflection (the {\small{PEXRIV}} model from
Magdziarz \& Zdziarski 1995) coupled with a smeared edge and a
diskline to account for the drop at 7.4 keV and for the broad 5-7 keV
residuals, respectively. Even by adding the smeared edge component,
the reflection fraction is found to be large (R around 2, depending on
the continuum parametrisation).

\subsection*{A possible interpretation}
\begin{figure}
\rotatebox{270}{
\resizebox{!}{\columnwidth}
{\includegraphics{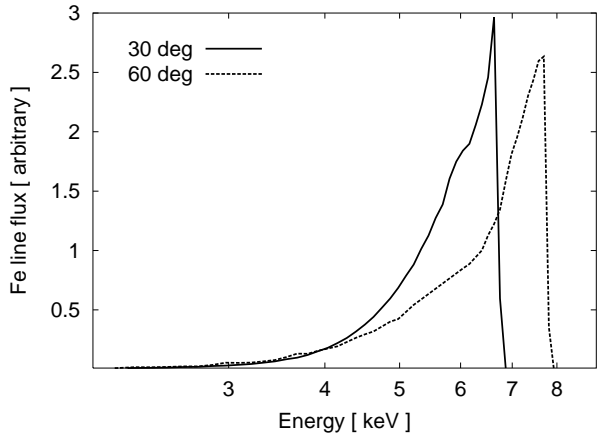}}}
%{\includegraphics{degrees.eps}}}
%\vspace{0.3cm}
\caption{We show, as an example, the dependency of the iron line
  profile from the observer inclination. The line profiles are
  obtained with the same source configuration, have both a rest frame
  energy of 6.4~keV and only differ by observer inclination. The blue wing
  of a neutral, 6.4~keV line could in principle account for the
  spectral drop at 8~keV seen in IRAS~13224--3809 is seen at
  sufficiently high inclination. Larger inclinations would produce a
  blue wing at even higher energy than shown. A higher line energy
  (due to ionisation) would reduce the constraint on the inclination
  angle, but is not necessarily required to explain the drop.}
\label{degrees}
\end{figure}

We have seen that the spectra of some NLS1 galaxies can be interpreted
as reflection--dominated with strong and broad iron line features,
although different models such as partial--covering can describe the
data equally well; only future, better data will allow, in our
opinion, to discriminate between the different possible
interpretations.

The light bending model can produce reflection--dominated spectra by
simply varying one parameter (the primary source height) toward its
extreme value ($h_s \rightarrow 0$) so that a transition into regime I
is produced. These reflection--dominated spectra would be naturally
associated with a strong iron emission line with different possible
degrees of ionisation. Large values of the EW, such as those needed to
interpret the drops at 7 and 8 keV in the spectra of 1H~0707--495 and
IRAS~13224--3809 as the blue wing of a relativistic line, can in
principle be reached if a certain degree of ionisation is allowed
\cite{gmfr93,gmfr96}, as expected if the source height is low and
illumination of the inner disc very strong.

We point out here that in order to produce a drop at 7--8~keV no
extreme ionisation of the line is needed.  Even a neutral 6.4~keV line
can have a blue wing at 8~keV if seen at sufficiently high
inclination. As an example, in Fig. \ref{degrees} we show how neutral
(i.e.  6.4~keV~) iron lines are observed at inclinations of 30 and 60
degrees. It is clear from the Figure that the blue wing of an iron
line seen at sufficiently high inclination can be located around
8~keV even if ionisation is not considered. Notice that a high
inclination has sometimes been invoked to account for both the strong
soft excess and the giant variability of this source (e.g. Boller et
al. 1997). A higher line energy due to ionisation would allow a lower
inclination of the system but is not formally required to explain the
8~keV drop of IRAS~13224--3809, while, as mentioned, ionisation could
help to increase the line EW.

Moreover, a very good description of the 1H~0707--495 data can be
obtained by assuming that the spectrum is reflection--dominated as a
result of multiple reflections from different layers of material.
This situation may be expected, for example, if the system is
characterised by a high accretion rate and disc instabilities are
strong enough to produce large density inhomogeneities in the disc. A
multiple reflection model gives an extremely good description
($\chi^2_\nu = 287 / 303$) to the data and is able to reproduce the
drop at 7~keV \cite{fbetal02,rfb02}. Within this model, the large
value of the line EW is produced self--consistently together with the
reflection continuum.

Multiple reflection models have not yet been tested on
IRAS~13224--3809, while a reflection--dominated spectrum due to 
multiple reflection (from the ionised reflection models of Ross
\& Fabian 1993 and Ballantyne, Iwasawa \& Fabian 2001) gives a
reasonable fit to the broadband spectrum of IRAS~13349+2438
\cite{letal03}. Furthermore, this model self--consistently accounts
for the edge and the line without the necessity to invoke
phenomenological models (such as the smeared edge, see discussion
above).

Gravitational light bending could also give rise to multiple
reflections: the reflected spectrum is partly forced to return to the
disc by the strong gravitational field of the central black hole. In
this case the spectrum produced by the second reflection would be
predominantly emitted in the innermost regions of the disc, where
lensing is stronger and thus most of the returning radiation is
concentrated. The effect of the returning radiation (from the disc to
the disc itself) would produce a steeper emissivity profile in the
inner regions of the accretion disc than in the outer and a further
reddening/broadening of the iron line (and of the RDC). 

However, the overall effect of multiple reflections on the spectrum is
complex: the spectra due to multiple reflections qualitatively
resemble those due to single reflection with slightly higher
ionisation parameters and much higher iron abundances. In particular,
the relative strength of the iron emission line is enhanced with
respect to the case of a single reflection and, therefore, large
values of the line EW such as those needed to explain the drops in
1H~0707--495 and IRAS~13224--3809 in terms of reflection can be
obtained (Ross, Fabian \& Ballantyne 2002).

\section{Applications to Galactic Black Hole Candidates}

Our model for the spectral variability of AGN can, in principle, be
applied to Galactic Black Hole candidates (BHC) as well. Strong, broad
iron lines have been found in several BHC (e.g.  Martocchia et al
2002; Miller et al 2002a,b,c and 2003a,b; Miniutti, Fabian \& Miller
2003), some of which suggest rapid spin. The lines are often seen when
the source is in a very high or intermediate state. In particular, in
the cases of GX~339-4 and XTE~J1650--500, the broad, relativistic iron
line is associated with a very steep emissivity profile on the
accretion disc and a very small inner disc radius, suggesting that the
illuminating source is centrally concentrated and that the central
black hole is rapidly spinning\footnote{Or, if the black hole is
  slowly/non rotating, that the contribution from the plunging region
  is not negligible.}, as we are assuming in this work. Thus, these
X--ray sources might represent laboratories to test our model in
stellar mass black hole accreting systems.

A recent analysis based on {\it{BeppoSAX}} data of the XTE~J1650--500
very high state evolution during its 2001 outburst, reveals that the
(broad) iron line variability with respect to the observed PLC matches
well the predictions of the light bending model \cite{mfm03}. The same
qualitative behaviour is also found in the long--term analysis of the
{\it{RXTE}} data by Rossi et al. (2003; 2004, in preparation). The PLC
decreases by about one order of magnitude during the outburst
evolution, while the iron line drops only by a factor 3 following,
with some scatter, the behaviour shown in Fig \ref{ff} during regimes
I and II. The scatter could be introduced by the (here not modelled)
variation in the power law photon index (that affects the iron line
flux because the spectral shape of the illuminating continuum on the
disc changes), or by intrinsic luminosity variations of the primary
source.  Moreover, the spectrum of XTE~J1650--500 in the
{\it{BeppoSAX}} data appears to become more and more
reflection--dominated as the PLC decreases, again in excellent
agreement with the predictions of the light bending model.

We also note that the iron line during the 2002 outburst of
4U~1543--47 shows a similar behaviour and seems to follow the path
shown in Fig.\ref{ff} (with some scatter and exceptions that could be
due also to the observed intermittent radio activity during
the outburst; see Park et al. 2003).

It would be remarkable if a simple model, such as the one we are
proposing here, could account for some (if not most) of
the spectral variability in some AGN as well as in a Galactic BHC such
as XTE~J1650--500 during its very high state evolution. Further
observational study of the evolution of power law and thermal
components and, if present, RDC and broad iron lines in BHC is needed
to assess the relevance of light bending in these sources and to
disentangle relativistic effects from different, obviously present,
physical mechanisms. Due to the similarities between XTE~J1650--500
and GX~339--4, we suggest that the very high state evolution
in these two BHC could be similar and characterised by a drop of the PLC
associated with an increase of the reflection fraction and with broad
iron line variability according approximately to Fig. \ref{ff}.

It is not our purpose to explain the behaviour of BHC here, but merely
note that some of the observed variability may be accounted for by the
strong light bending effects discussed here. Some of the extreme
variability (see again, e.g.  XTE~J1650--500; Tomsick et al. 2003)
could for example also be due to light bending effects that can occur
on very short timescales in low--mass systems (see Section 3.3). 

We point out here that broad iron lines are seen in AGN such as
NGC~4051 and MCG--6-30-15 whose power spectral density (PSD) shows
remarkable common features with that of the BHC Cyg~X-1 in its high
rather than low state \cite{vfn03,mark03,mac03}. Furthermore a power
law spectral component is seen most of the time in the high state of
Cyg~X-1, making this state more similar to the very high state of
other BHC, where both a thermal and a power law component are present
(see M$^{\rm{c}}$Clintock \& Remillard 2003; Pottschimdt et al. 2003).
Broad iron lines in BHC, as mentioned, are often observed in
intermediate (e.g. Miller et al.  2002c) or in very high states (e.g.
Miller et al. 2002a, Miniutti, Fabian \& Miller 2003).  We support
therefore previous claims (see e.g.  M$^{\rm{c}}$Hardy et al. 2003) of
a similarity between BHC in these states and AGN such as NGC~4051 and
MCG--6-30-15, suggesting a possible geometrical origin for such a
similarity (i.e. the presence of a centrally concentrated primary
source close to the central black hole).

\section{Discussion and conclusions}

We have explored a light bending model \cite{fv03,mfgl03} in which the
observed flux is strongly correlated with the height ($h_s$) of the
primary X--ray source above the accretion disc around a massive black
hole. High observed fluxes correspond to large $h_s$ while if the
source height is small (few gravitational radii) light bending forces
most of the primary emission to be bent onto the disc dramatically
reducing the flux and enhancing the reflection features.
  
Even without invoking intrinsic luminosity variation (likely to be
present both on short and long timescales) changes of more than one
order of magnitude in the direct continuum flux can be produced by
light bending alone. Variability induced by light bending can act on
such short timescales that even the large and fast variability of some
NLS1 galaxies (e.g. IRAS~13224--3809) could be explained without
breaking the ``efficiency limit''.

The reflection--dominated component (RDC) is variable with much
smaller amplitude than the direct continuum (PLC) and three different
regimes can be identified in which the RDC is correlated,
anti--correlated or almost independent with respect to the direct
continuum. These regimes correspond to low, high, and intermediate flux
states respectively.  As a general rule, the iron line EW is
anti--correlated with the continuum and becomes almost constant only
during very low or very high flux states.

At low fluxes (low source height) the spectrum becomes naturally
reflection--dominated because of light bending so that large values of
the reflection fraction are expected during particularly low flux
states. Furthermore the reflected radiation as well can only partly
escape the strong gravitational field of the central black hole and
some returns to the disc producing multiple reflection from the inner
regions of the disc and reddening the observed spectrum. Multiple
reflections could also account for large values of the iron line EW,
because the overall spectra strongly resemble those due to single
reflection but with much higher iron abundance.

Low flux states are expected to be associated with broad iron emission
lines.  However, it is possible that the line is in some cases so
broad (and weak) that chances of detection with current instruments
are seriously compromised.  Moreover, the ionisation of the inner
regions of the disc (likely if the source height is small) may
contribute to reduce the red wing of the line and to lower further the
chances of detection.

The emission from the disc (and from the primary source) is also
strongly beamed in the equatorial plane due to gravitational lensing
and special relativistic beaming. One thus expects an additional
reflection component from the outer regions of the disc (if the disc
is flared and diverge from the equatorial plane) and/or the putative
molecular torus. Both components should produce narrow features in the
spectra that would be enhanced during low flux states, although
depending on the particular source geometry. We thus predict a general
anti--correlation between the EW of the narrow lines and X--ray flux
(that could be related to the observed X--ray Baldwin effect). The
beaming of the emission along the equatorial plane may be relevant
also to the properties of the narrow optical emission lines in NLS1
galaxies.  Due to the anisotropy of the illumination, the BLR appear
as if they were flattened in the equatorial plane. An observer at low
or moderate inclination would then detect narrow optical emission
lines.
  
The light bending model proves able to account for many of the
puzzling properties of MCG--6-30-15. In particular, if one
interprets the low flux state of MCG--6-30-15 as regime I, the normal
state as regime II and the high state as regime III, the following
observational facts are fully recovered by our model (see also
Miniutti et al. 2003):
\begin{enumerate}
\item the timescale for the variability of MCG--6-30-15 is totally
  consistent with the capabilities of the model;
\item in normal flux states, MCG--6-30-15 is well
  described by a two--component model comprising a variable PLC
  (within a factor $\sim$~4) and an almost constant RDC and iron line
  (regime II);
\item the iron line profile of MCG--6-30-15 in normal flux states is
  well represented by a typical profile from regime
  II. The emissivity of the disc is in the form of a broken power law,
  steeper in the inner disc than in the outer;
\item the iron line EW in normal flux states is anti--correlated with
  the continuum, as we predict for regime II;
\item in low flux states, the line appears to be broader, correlated
  with the continuum and an almost constant EW is observed (regime I);
\item the spectrum of MCG--6-30-15 appears to be more and more
  reflection--dominated as the flux decreases
\item in high flux states (regime III or II/III), the iron line
  appears to be generally narrower than in lower flux states (regime
  I or I/II).
\end{enumerate}
All these properties are very well reproduced by our model and
represent different, and mostly observationally independent, tests for
any theoretical attempt to explain the MCG--6-30-15 spectrum and
variability. The consistent behaviour of line flux, line EW, line
profile, and relative strength of disc reflection with respect to the
PLC variations strongly suggests that general relativistic effects in
the near vicinity of the central (likely rotating) black hole provide
an important contribution to the variability and that light bending is
a relevant aspect of the overall picture.

Moreover, the NLS1 galaxy NGC~4051 exhibits spectral and variability
properties that could also be explained by gravitational bending of
the primary X--ray emission, even in the most extreme cases such as
the ultra--dim and completely reflection--dominated spectrum observed
in 1998. Possibly more remarkable (for the different mass scale with
respect to MCG--6-30-15) is the behaviour of the Galactic black hole
candidate XTE~J1650--500 whose broad, relativistic iron line exhibits
qualitatively the same variability behaviour as predicted by the light
bending model during regimes I and II. These results suggest that the
model we are proposing may well act in both supermassive and
stellar mass black holes.

The variability of some black hole systems (in some epochs, such as
the very high state, for BHC) could then be dominated by changes in
the primary source height resulting in an average positive correlation
between the RDC (and iron line) and the continuum if the source is in
a low flux state (regime I or I/II), in an almost constant RDC/line if
the source is in an intermediate flux state (regime II) and in
anti--correlation in high flux states (regime II/III or III).
Additional variability is likely to be provided by intrinsic
luminosity variations, changes in the motion of the primary source
and/or the presence of multiple emitting regions that are active at
different times and locations (or heights). The ionisation
structure of the surface of the accretion disc, which we have
neglected for simplicity, can also be relevant and slightly modify our
predictions.

However, we have shown that relativistic effects such as light bending
can not be neglected, especially if some indication of a centrally
concentrated primary source is found in observations (e.g. a steep
emissivity profile, a broad iron line ...). The location
($\rho_s\,,\,h_s$) of the primary source represents an additional
parameter that should be included in the analysis of the spectral
variability. The light bending model provides a simple explanation for
the properties of some X--ray sources and may well produce the bulk of
the observed variability reconciling, in many cases, observations with
theoretical models for reflection from accretion discs. It is also
clear that a combination of light bending and intrinsic luminosity
variations would cover a larger parameter space and could be relevant
to interpret the spectral variability of many more sources than
discussed in this work.

\section*{Acknowledgements}
We thank Russell Goyder and Anthony Lasenby for help with the initial
numerical code that was used to perform some of the computations. We
are grateful to Kazushi Iwasawa, Giorgio Matt, Phil Uttley and Simon
Vaughan for constructive comments and discussions, and to the
anonymous referee for suggestions and criticisms which helped to
improve our work. We thank the Institute of Astronomy for the use of a
34-processor machine. GM thanks the PPARC for support. ACF thanks the
Royal Society for support.

\end{document}